\documentclass[twocolumn]{aastex63}

\usepackage{ulem}
\usepackage{amsmath}
\received{October 8, 2021}
\revised{November 3, 2021}
\accepted{November 10, 2021}
\submitjournal{Astrophysical Journal Letters}

\shorttitle{Molecular-Gas to Dust Mass Ratios in Quiescent Galaxies}
\shortauthors{K. Whitaker, D. Narayanan, et al.}

\begin{document}

\title{High Molecular-Gas to Dust Mass Ratios Predicted in Most Quiescent Galaxies}

\correspondingauthor{Katherine E. Whitaker}
\email{kwhitaker@astro.umass.edu}

\author[0000-0001-7160-3632]{Katherine E. Whitaker}
\affil{Department of Astronomy, University of Massachusetts, Amherst, MA 01003, USA}
\affil{Cosmic Dawn Center (DAWN), Denmark}

\author[0000-0002-7064-4309]{Desika Narayanan}
\affil{Department of Astronomy, University of Florida, 211 Bryant Space Sciences Center, Gainesville, FL 32611 USA}
\affil{Cosmic Dawn Center (DAWN), Denmark}

\author[0000-0003-2919-7495]{Christina C. Williams}
\affil{Steward Observatory, University of Arizona, 933 North Cherry Avenue, Tucson, AZ 85721, USA}

\author[0000-0001-8015-2298]{Qi Li}
\affil{Max-Planck-Institut f{\"u}r Astrophysik, Karl-Schwarzschild-Str. 1, Garching b. M{\"u}nchen D-85741, Germany}
\affil{Department of Astronomy, University of Florida, 211 Bryant Space Sciences Center, Gainesville, FL 32611 USA}

\author[0000-0003-3256-5615]{Justin S. Spilker}
\altaffiliation{NHFP Hubble Fellow}
\affiliation{Department of Astronomy, University of Texas at Austin, 2515 Speedway, Stop C1400, Austin, TX 78712, USA}
\affiliation{Department of Physics and Astronomy and George P. and Cynthia Woods Mitchell Institute for Fundamental Physics and Astronomy, Texas A\&M University, 4242 TAMU, College Station, TX 77843-4242}

\author[0000-0003-2842-9434]{Romeel Dav\'{e}}
\affil{Institute for Astronomy, Royal Observatory, Univ. of Edinburgh, Edinburgh EH9 3HJ, UK}
\affil{Dept. of Physics and Astronomy, University of the Western Cape, Bellville, Cape Town 7535, South Africa}
\affil{South African Astronomical Observatories, Observatory, Cape Town 7925, South Africa}

\author[0000-0002-3240-7660]{Mohammad Akhshik}
\affil{Department of Physics, University of Connecticut,
Storrs, CT 06269, USA}

\author[0000-0003-3596-8794]{Hollis B. Akins}
\affil{Department of Physics, Grinnell College, 1116 Eighth Ave., Grinnell, IA 50112, USA}

\author[0000-0001-5063-8254]{Rachel Bezanson}
\affil{Department of Physics and Astronomy, University of Pittsburgh, Pittsburgh, PA 15260, USA}

\author[0000-0002-3097-5381]{Neal Katz}
\affil{Department of Astronomy, University of Massachusetts, Amherst, MA 01003, USA}

\author[0000-0001-6755-1315]{Joel Leja}
\affil{Department of Physics, 104 Davey Lab, The Pennsylvania State University, University Park, PA 16802, USA}
\affil{Institute for Computational \& Data Sciences, The Pennsylvania State University, University Park, PA, USA}
\affil{Institute for Gravitation and the Cosmos, The Pennsylvania State University, University Park, PA 16802, USA}

\author[0000-0002-4872-2294]{Georgios E. Magdis}
\affil{Cosmic Dawn Center (DAWN), Denmark}
\affil{DTU-Space, Technical University of Denmark, Elektrovej 327, 2800, Kgs. Lyngby, Denmark}
\affil{Niels Bohr Institute, University of Copenhagen, Lyngbyvej 2, DK-2100 Copenhagen, Denmark}

\author[0000-0002-8530-9765]{Lamiya Mowla}
\affil{Dunlap Institute for Astronomy and Astrophysics, University of Toronto, 50 St George St, Toronto, ON M5S 3H4, Canada}

\author[0000-0002-7524-374X]{Erica J. Nelson}
\affil{Department of Astrophysical and Planetary Sciences, 391 UCB, University of Colorado, Boulder, CO 80309-0391, USA}

\author[0000-0001-8592-2706]{Alexandra Pope}
\affil{Department of Astronomy, University of Massachusetts, Amherst, MA 01003, USA}

\author[0000-0003-3474-1125]{George C. Privon}
\affil{National Radio Astronomy Observatory, 520 Edgemont Rd, Charlottesville, VA 22903, USA}

\author[0000-0003-3631-7176]{Sune Toft}
\affil{Cosmic Dawn Center (DAWN), Denmark}
\affil{Niels Bohr Institute, University of Copenhagen, Lyngbyvej 2, DK-2100 Copenhagen, Denmark}

\author[0000-0001-6477-4011]{Francesco Valentino}
\affil{Cosmic Dawn Center (DAWN), Denmark}
\affil{Niels Bohr Institute, University of Copenhagen, Lyngbyvej 2, DK-2100 Copenhagen, Denmark}

\begin{abstract}  
Observations of cold molecular gas reservoirs are critical for understanding the shutdown of star formation in massive galaxies. While dust continuum is an efficient and affordable tracer, this method relies upon the assumption of a ``normal'' molecular-gas to dust mass ratio, $\delta_{\mathrm{GDR}}$, typically of order one hundred. Recent null detections of quiescent galaxies in deep dust continuum observations support a picture where the cold gas and dust has been rapidly depleted or expelled.  In this work, we present another viable explanation: a significant fraction of galaxies with low star formation per unit stellar mass are predicted to have extreme $\delta_{\mathrm{GDR}}$ ratios.  We show that simulated massive quiescent galaxies at $0 < z < 3$ in the \textsc{simba} cosmological simulations have $\delta_{\mathrm{GDR}}$ values that extend $>$4 orders of magnitude. The dust in most simulated quiescent galaxies is destroyed significantly more rapidly than the molecular gas depletes, and cannot be replenished. The transition from star-forming to quiescent halts dust formation via star formation processes, with dust subsequently destroyed by supernova shocks and thermal sputtering of dust grains embedded in hot plasma. After this point, the dust growth rate in the models is not sufficient to overcome the loss of $>$3 orders of magnitude in dust mass to return to normal values of $\delta_{\mathrm{GDR}}$ despite having high metallicity. Our results indicate that it is not straight forward to use a single observational indicator to robustly pre-select exotic versus normal ratios. These simulations make strong predictions that can be tested with millimeter facilities.
\end{abstract}


\section{Introduction}
\label{sec:intro}

After over a decade and thousands of hours of observations, we now have a reasonably good 
census of the dust and cold gas content in ‘normal’ star forming galaxies out to $z\sim2$, as well
as other galaxy properties that correlate \citep[see reviews by, e.g.,][]{Carilli13, Hodge20, Tacconi20}. 
However, the dust and cold gas content of quiescent galaxies remain uncertain, especially towards
higher redshift.  Spectroscopic studies measuring the cold gas content of quiescent galaxies at $z>0.5$
are limited, as
robust constraints are observationally expensive.
First results from intermediate redshift studies 
using CO emission to trace metal-rich molecular hydrogen are conflicting: some yield surprisingly large gas reservoirs \citep[e.g.,][]{Suess17,Rudnick17,Hayashi18,Belli21}, while others 
result in strong upper limits suggesting rapid gas depletion \citep[e.g.,][]{Sargent15, Bezanson19, Williams21}, and one study results in a mixture \citep{Spilker18}.  The diversity in M$_{\mathrm{H2}}$ may be due to the range in specific star formation rates (sSFR$\equiv$SFR/M$_{\star}$) of the samples themselves, as low sSFRs are hard to constrain \citep[e.g.,][]{Leja19}.

By stacking low-resolution far-infrared to sub-millimeter imaging, the first studies of statistically meaningful 
samples of quiescent galaxies at $z\sim1-2$ find moderate dust and inferred gas content of order $f_{\mathrm{H2}}= \mathrm{M}_{\mathrm{H2}}/\mathrm{M}_{\star}\sim5-10\%$ \citep{Gobat18,Magdis21}.  These results are within 2$\sigma$ of an extrapolation of the \citet{Tacconi18} scaling relations of sSFR and $f_{\mathrm{H2}}$. 
However, deblending low-resolution data is non-trivial and systematic uncertainties can be large \citep{Viero12}.  
While these analyses perform a comprehensive modeling of the full infrared spectral energy distribution (SED), local studies often leverage correlations between the total gas mass and a single far-infrared band. That said, debate remains regarding which wavelengths provide the strongest correlation.
Some studies find the highest correlation of the molecular-gas mass with dust mass as traced by
the longest wavelengths \citep[e.g.,][]{Bourne13}, whereas
others find that the molecular-gas mass is best correlated with the obscured SFR at the peak of the infrared SED \citep[rest-frame 100-160$\mu$m;][]{Groves15}.  Moreover, there is not yet consensus whether the gas traced 
by the dust continuum is primarily molecular or a combination with neutral hydrogen \citep{Janowiecki18}.  

\citet{Scoville16} calibrate the Rayleigh-Jeans dust continuum as a tracer of molecular-gas mass, using a sample of star-forming galaxies at $z\sim0$ and $z\sim2$ with both CO(1-0) and dust continuum fluxes.
Combining this methodology based on optically-thin dust continuum with sensitive millimeter telescopes may be one of the most efficient ways to observe 
gas in distant quiescent galaxies. Uncertainties due to variations in dust temperature are thought to be minimal at $\lambda_{\mathrm{rest}}>250$ $\mu$m \citep[e.g.][]{Scoville14}, though some tension exists at high redshift \citep{Harrington21}.  
The major driver of the scatter in the conversion of dust continuum to gas mass is instead the ratio of the molecular-gas mass to dust mass, $\delta_{\mathrm{GDR}}$=M$_{\mathrm{H2}}$/M$_{\mathrm{dust}}$ \citep{Privon18}. 
Dust continuum is demonstrated to be a robust tracer of the molecular-gas mass for `normal' star-forming galaxies \citep{Privon18, Liang18,  Kaasinen19}, but calibrations are not yet tested for quiescent galaxies.
Inherent to the \citet{Scoville16} methodology is the assumption that $\delta_{\mathrm{GDR}}$=150, though it is bundled together with the assumed dust emissivity per unit mass.
High $\delta_{\mathrm{GDR}}$ ratios in nearby quiescent galaxies are thought to be the result of thermal sputtering resulting from the impact of dust 
grains with hot plasma \citep[e.g.,][]{Galliano18, Smercina18}.  

\begin{figure*}[t]
\leavevmode
\centering
\includegraphics[width=\linewidth]{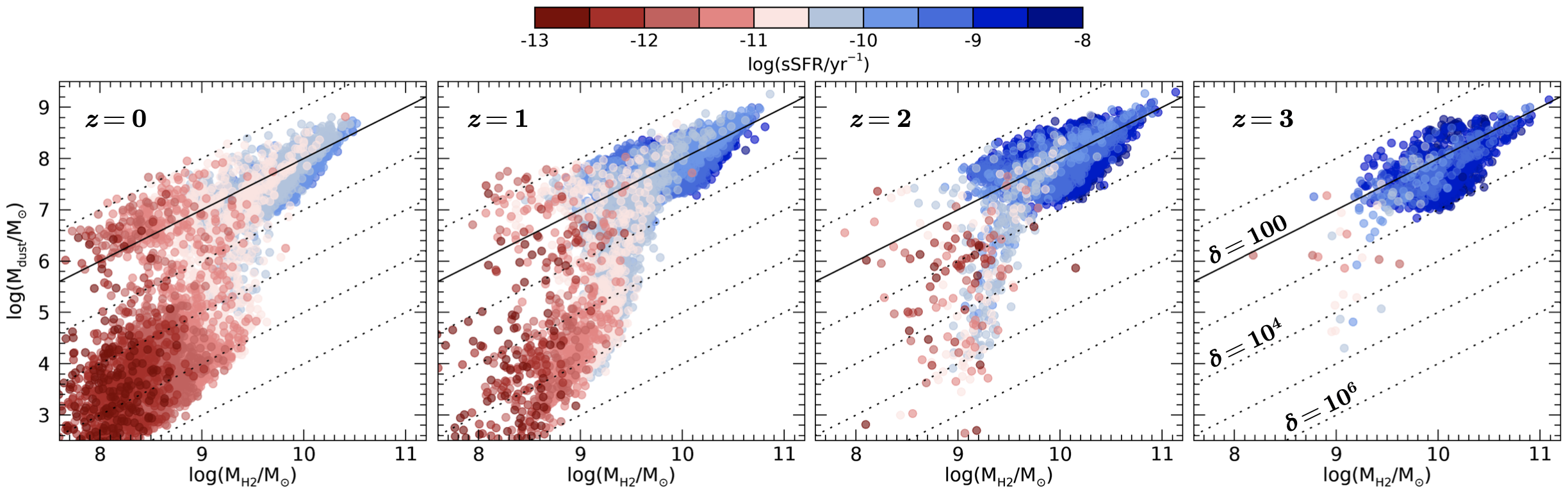}
\caption{Dust masses are dramatically lower than the molecular-gas mass for model galaxies with low sSFRs (red), assuming the standard relative abundance of $\delta_{\mathrm{GDR}}$=M$_{\mathrm{H2}}$/M$_{\mathrm{dust}}$=100 (solid line).  The trend becomes most pronounced by $z$=1, as the quiescent population grows. The models predict that dust is rapidly destroyed when star formation tapers.}
\label{fig:sample}
\end{figure*}

Even when leveraging strong gravitational lensing, \citet{Whitaker21} detect cold dust in only two out of a sample of six quiescent galaxies at $z\sim2$.  
When assuming $\delta_{\mathrm{GDR}}$ of 100, these data impose strong upper limits of $f_{\mathrm{H2}}<1\%$.
The most obvious conclusion from the null detections is that the galaxies harbor very small dust and molecular-gas reservoirs.  This 
is consistent with other studies using independent tracers of molecular hydrogen that find similarly low gas fractions among a sample of massive quiescent galaxies \citep[e.g.,][]{Williams21}.
However, sample sizes remain small and we cannot rule out that the assumed $\delta_{\mathrm{GDR}}$ ratio is actually much higher at low sSFR, implying under-predicted inferred gas fractions. 

In this paper, we seek to gain physical intuition
on $\delta_{\mathrm{GDR}}$ ratios at low sSFR by studying populations of simulated galaxies ranging from $z$=0 to $z$=3 in the \textsc{simba} galaxy formation simulation that separately tracks $H_2$ and dust. We present details of the simulated galaxy sample in \S\ref{sec:methods}, and show results in \S\ref{sec:mocks} including a comparison to the literature.  In \S\ref{sec:discussion}, we discuss the implications of a theoretical model tuned to reproduce dust-to-metal ratios at $z$=0 in the context of the existing astronomical literature.  \textsc{simba} adopts a $\Lambda$CDM cosmology with $\Omega_{M}$ = 0.3 and H$_0$ = 68 km s$^{-1}$ Mpc$^{-1}$, and a \citet{Chabrier} initial mass function.

\section{Numerical Methods}
\label{sec:methods}

We turn to hydrodynamic cosmological simulations that include predictive modeling of the gas and dust physics to better understand this problem.  
\citet{Li19} present a self-consistent model for the dust-to-gas and dust-to-metal ratios in galaxies in the \textsc{simba} cosmological hydrodynamic galaxy formation simulation \citep{Dave19}, tracking the production, growth, and destruction of dust grains over time, broadly following  \citet{McKinnon17} with notable improvements \citep{Li19}.
The dust content is primarily governed by (1) formation
of dust in stellar ejecta (e.g., Type II supernovae and asymptotic giant branch stars), 
(2) growth of dust via the accretion of metals, and (3) destruction
of dust via thermal sputtering, consumption by star formation, or supernovae (SNe) shocks.  
Molecular-gas is computed within dense ($n_H>0.13$~cm$^{-3}$) gas using a metallicity-dependent prescription following \citet{KrumholzGnedin2011}.
We refer to \cite{Dave19}, \citet{Li19}, and \citet{Narayanan20}
for full details on \textsc{Simba}.

We select a total of 17869 simulated galaxies having log(M$_{\star}$/M$_{\odot}$)$>$10, identified via a 6-D friends-of-friends algorithm applied to dense gas and stars in four snapshots ($z$=0,1,2,3) of the (100 $h^{-1}$ Mpc)$^3$ \textsc{simba} simulation. 
This \textsc{Simba} run has a mass resolution of 9.6$\times$10$^{7}$ M$_{\odot}$ for dark matter particles and 1.82$\times$10$^{7}$ M$_{\odot}$ for gas elements and stars. 
Tests at 8$\times$ higher mass resolution performed by \citet{Dave19} and \citet{Li19} find M$_{\mathrm{dust}}$ and $\delta_{\mathrm{GDR}}$ converge.
The sample comprises 7449 (6337) galaxies at $z$=0 above (below) log(sSFR)$=-10$ yr$^{-1}$, 6691 (2837) at $z$=1, 2713 (353) at $z$=2, and 1016 (36) at $z$=3.   
24\%, 7\%, and 2\% of the sample at $z=0$, 1, and 2, respectively, have no $H_2$ or dust; we discard these since we cannot compute $\delta_{GDR}$.
7\% of the remaining galaxies at $z$=0 have an instantaneous SFR=0 ($<$1\% at higher redshifts), which we arbitrarily set to 0.01 M$_{\odot}$ yr$^{-1}$ for numerical convenience; this has no impact on our results.   

Herein, we define $f_{\mathrm{H2}}$=M$_{\mathrm{H2}}$/M$_{\star}$ from the simulations and ignore neutral hydrogen.
The \citet{Scoville16} methodology and CO measurements \citep{Bolatto13} are both calibrated to 
trace molecular hydrogen alone, under the assumption that the fraction of dust associated with molecular hydrogen is greater than that of neutral hydrogen.  
This is not necessarily true for quiescent galaxies; if dust traces both molecular and neutral hydrogen, with significant neutral reservoirs expected \citep[e.g.,][]{Zhang19}, dust continuum serves as upper limits for M$_{H2}$. However, defining the gas mass to include neutral hydrogen (i.e., M$_{\mathrm{gas}}$=M$_{HI}$+M$_{H2}$) only exaggerates the implied exotic values of $\delta_{\mathrm{GDR}}$.  
Thus, the molecular hydrogen is summed for all particles within 30 kpc physical spherical aperture. Dust is fully coupled to the gas and its mass is calculated in the same way.  This physical aperture is important for removing contributions from nearby small, unresolved galaxies, which can be significant for a quenched galaxy that has very little molecular-gas itself.

\begin{figure}[t]
\leavevmode
\centering
\includegraphics[width=\linewidth]{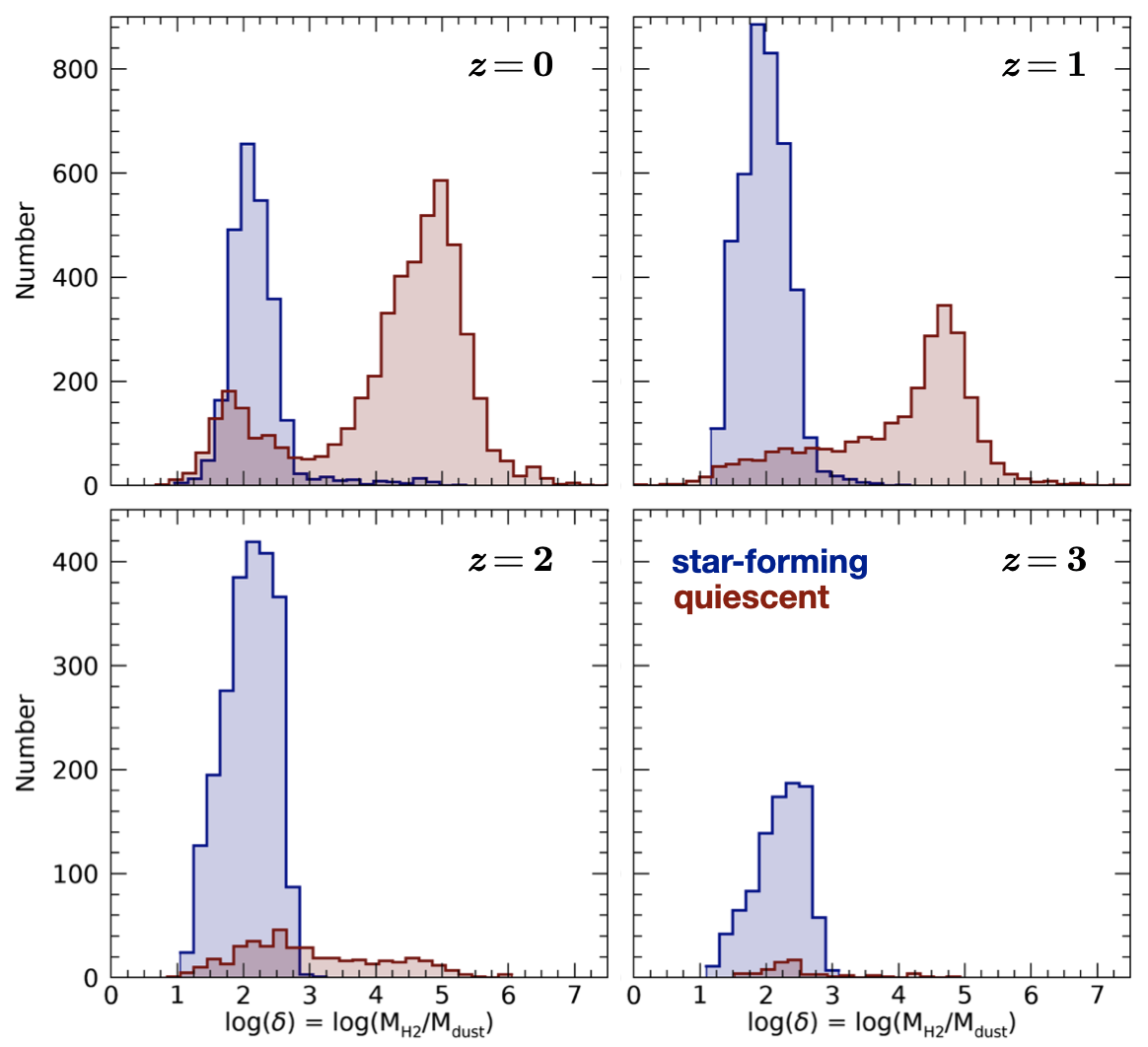}
\caption{Histograms of the molecular-gas to dust ratio, $\delta_{\mathrm{GDR}}$, for \textsc{simba} model galaxies from $z$=0 (top left) to $z$=3 (bottom right).  Star-forming galaxies peak at $\delta_{\mathrm{GDR}}\sim100$, whereas we see the buildup of the quiescent population spanning $\sim$4 orders of magnitude and peaking at $\delta_{\mathrm{GDR}}\sim10^{4}-10^{5}$. Once the dust is destroyed, rarely can a galaxy regrow it.  Galaxy populations are separated into star-forming and quiescent based on their distance from the log(SFR)-log(M$_{\star}$) relation.}
\label{fig:hist}
\end{figure}

\section{Molecular-gas to Dust Mass Ratios in the \textsc{simba} Cosmological Simulation}
\label{sec:mocks}

The relation between the molecular-gas mass and dust mass of \textsc{simba} model galaxy populations ranging from $z$=0 (left) to $z$=3 (right) is found in Figure~\ref{fig:sample}.  The solid line is the standard value of $\delta_{\mathrm{GDR}}$=100, with dotted lines offset in increments of 1 dex.
Quiescent galaxies are broadly identifiable on the red end of 
the color spectrum.  We show that star-forming galaxies (blue points) follow a relatively tight 
relation between dust mass and molecular-gas mass consistent with $\delta_{\mathrm{GDR}} \sim 100$,
but we see a large intrinsic scatter among the low sSFR population.  In particular, $\delta_{\mathrm{GDR}}$ 
for low sSFR galaxies spans six orders of magnitude and most objects have significantly higher ratios than standard assumptions of $\delta_{\mathrm{GDR}}\sim$100--200 \citep[e.g.,][]{Tacconi20}. 
This dramatic change in $\delta_{\mathrm{GDR}}$ is the result of rapidly decreasing dust masses.

\begin{figure*}[t]
\centering
\includegraphics[width=0.9\linewidth]{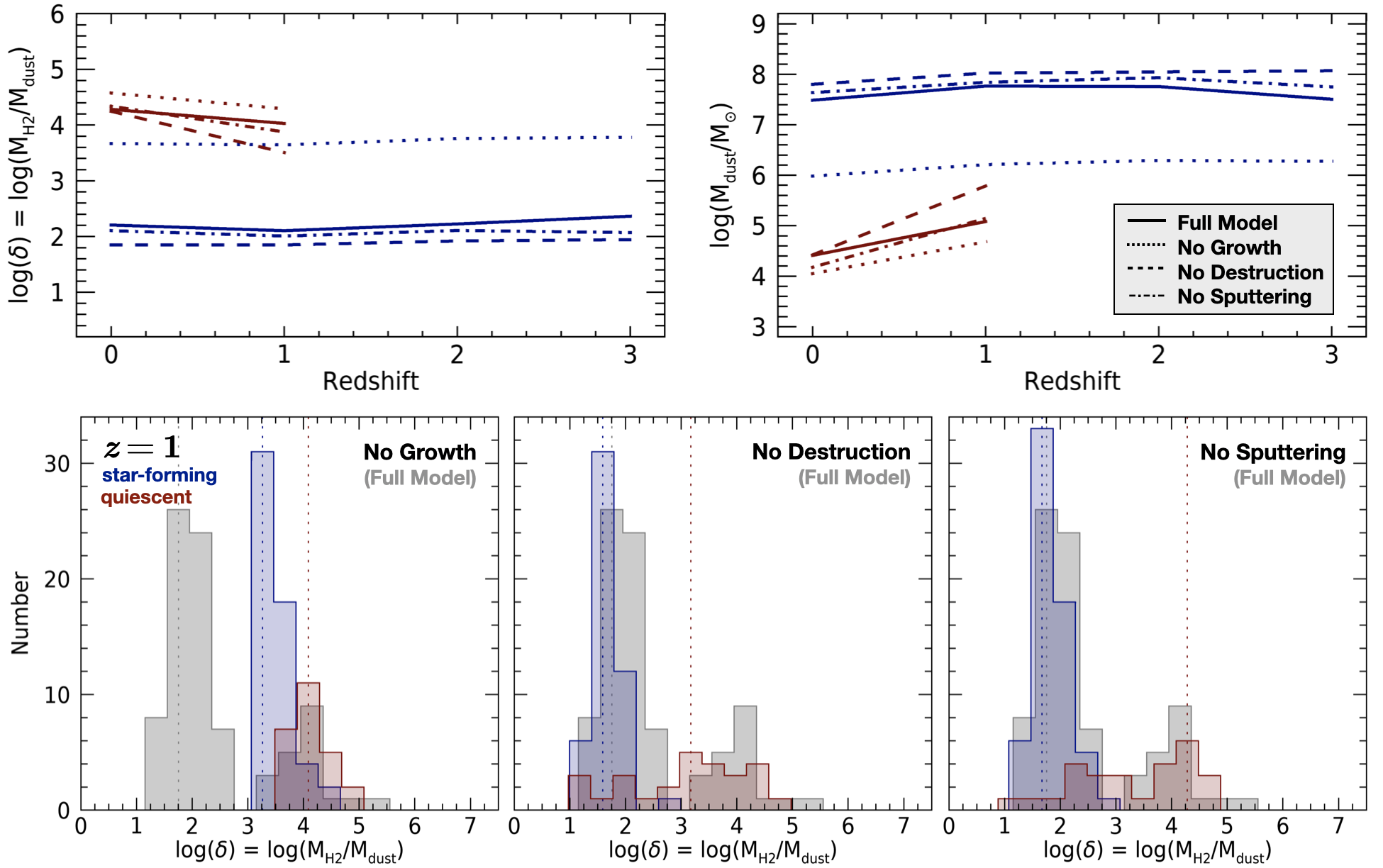}
\caption{Comparison of (25/h Mpc)$^{3}$ box simulation runs for the full model relative to (a) no dust growth, (b) turning off dust destruction (labeled ‘No Destruction’) by SN shocks and thermal sputtering, but not astration -- this is not possible to turn off, and (c) no thermal sputtering.  The top panels demonstrate the redshift evolution of the median molecular-gas to dust mass ratio (left) and dust mass (right) for star-forming (blue) and quiescent (red) galaxies. Only bins with greater than 10 galaxies are shown. The bottom panels show a comparison of the full model distribution at $z=1$ (gray, no separation of quiescent/star-forming) relative to each of the cases separated into quiescent/star-forming populations. Dotted lines represent the mode.}
\label{fig:test}
\end{figure*}

Figure~\ref{fig:hist} shows the redshift evolution of $\delta_{\mathrm{GDR}}$ when separating star-forming (blue) and quiescent (red) galaxies into bimodal galaxy populations.
In order to do this in a meaningful way, we match the log(SFR)-log(M$_{\star}$) relations parameterized by \citet{Whitaker14b} to the mean ridge-line of the model galaxies.  The polynomial fit to the observations at $z$=1-1.5, $z$=1.5-2 and $z$=2-2.5 are matched to running mean of the models at $z$=1, 2, and 3, respectively.  Shifting the $z$=0.5-1 fit down by 0.5 dex aligns with the model relation at $z$=0. These shifts remove known offsets between observed UV+IR SFRs and simulated datasets \citep{Nelson21}.  Our goal is to identify galaxies at a given epoch that are $>$0.7 dex below the star-forming sequence, labeling this population as quiescent.

Through this exercise, we learn something important: star-forming galaxies consistently peak at $\delta_{\mathrm{GDR}}\sim100$ with $\sigma$=0.4 dex, whereas the quiescent population has a wider range of $\delta_{\mathrm{GDR}}$.  This conclusion is robust to perturbations of the star-forming/quiescent boundary. In the models, the dust mass is governed by dust formation (Type II SNe, AGB stars), dust growth (accretion of metals), and dust destruction by thermal sputtering and SNe shocks.  
Here, thermal sputtering in the hot ISM and SNe shocks both contribute to the dust destruction process\footnote{This said, we should note that we do not model dust shattering in {\sc simba}, which may be relatively important \citep{Li21}.}.  To verify this, we conduct a series of controlled numerical experiments in which we run comparable  resolution (but smaller box) (25/h Mpc)$^{3}$ simulations with 256$^{3}$ particles, and systematically turn off dust growth, all destruction processes (except by star formation itself  -- this is not posssible as the dust is tied to gas particles in {\sc simba}), and just thermal sputtering (see Figure~\ref{fig:test}).
If we turn off dust growth altogether in the models, this has the most dramatic effect on the peak location of $\delta_{\mathrm{GDR}}$ for star-forming galaxies, shifting 2 dex higher.  Turning off dust destruction mechanisms does not completely remove the bimodality, indicating that dust destruction in star-forming regions also plays an important role. A key aspect here is that once the dust is destroyed, rarely can the galaxy regrow it, as indicated by the build-up of the quiescent population at $\delta_{\mathrm{GDR}}\sim10^4-10^5$. This bimodality within the quiescent population at $z=0$ is partially driven by the star-forming/quiescent definition, where some ambiguity exists.

\begin{figure*}[t]
\leavevmode
\centering
\includegraphics[width=\linewidth]{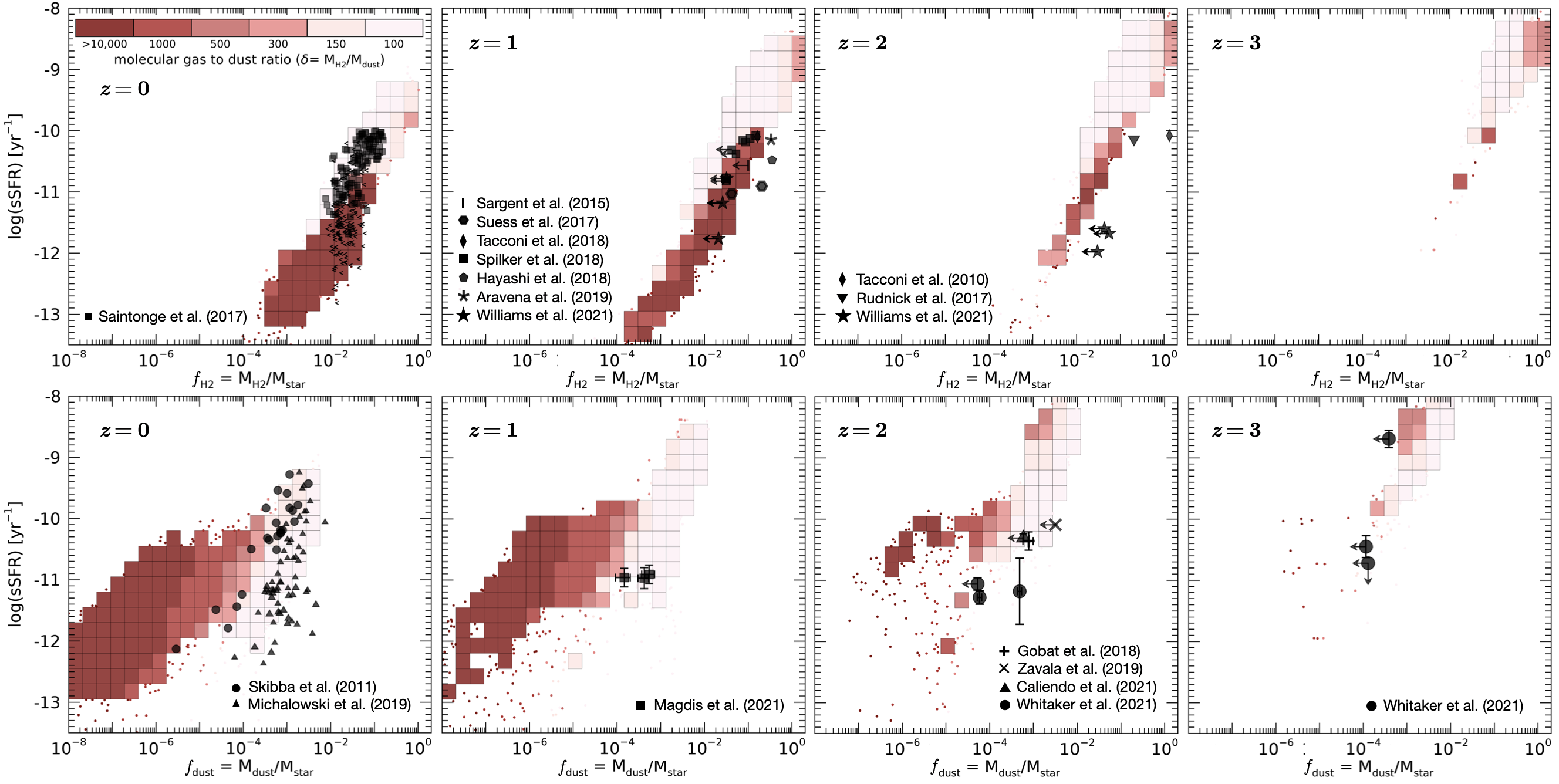}
\caption{\textbf{Top:} The models predict a tight correlation between decreasing $f_{\mathrm{H2}}$ with decreased sSFR. The squares represent the median molecular-gas to dust ratio, with small circles representing individual model galaxies plotted in sparsely populated parameter space. Black symbols represent CO measurements of quiescent galaxies in the literature, with upper limits indicated with an arrow (or left caret at $z$=0).  \textbf{Bottom:} $f_{\mathrm{dust}}$ fans out to cover up to 5 orders of magnitude below log(sSFR)$<$-10 yr$^{-1}$, given the predicted high molecular-gas to dust mass ratios.  Black symbols represent dust continuum measurements of quiescent galaxies in the literature.}
\label{fig:delta}
\end{figure*}

We color-code model galaxies by the median $\delta_{\mathrm{GDR}}$ value in Figure~\ref{fig:delta} to demonstrate the impact of the large intrinsic scatter in the factor transcribing $f_{\mathrm{dust}}$ (bottom) into $f_{\mathrm{H2}}$ (top).
This wide range in $\delta_{\mathrm{GDR}}$ produces a moderately tight relation between $f_{\mathrm{H2}}$ and log(sSFR), despite the variations observed in $f_{\mathrm{dust}}$.  Again, this is because the intrinsic scatter of $\delta_{\mathrm{GDR}}$ is driven by variations in M$_{\mathrm{dust}}$.  Figure~\ref{fig:delta} includes a compilation of CO \citep[top:][]{Saintonge17,Sargent15,Suess17,Tacconi18,Spilker18,Hayashi18,Aravena19,Williams21,Tacconi10,Rudnick17} and dust continuum observations \citep[bottom:][]{Skibba11,Michalowski19,Magdis21,Gobat18,Zavala19,Caliendo21,Whitaker21}. 

The lack of dust continuum detections at extremely low dust fractions may not be physical, but rather a selection bias. 
While the local early-type galaxies and red spirals from \citet{Rowlands12} \citep[compilation adopted from][]{Michalowski19} shown in Figure~\ref{fig:delta}
appear to have significantly larger dust masses than predicted by the models, this sample is blindly selected at sub-millimeter wavelengths and thus may not be representative.  Values of $f_{\mathrm{dust}}$ as low as 10$^{-5}$--10$^{-6}$ have been observed in nearby early-type galaxies \citep{Smith12}.  
However, placing such low limits on dust mass becomes technically challenging, even in the nearby universe.

If we take these predicted exotic molecular-gas to dust mass ratios at face value,
even with extremely sensitive limits of millimeter observations, there may be no hope to detect the lowest sSFR galaxies in dust continuum in the near term.  In some simulated metal-rich quiescent galaxies, dust regrowth is not enough to significantly increase the dust mass to counteract the destruction processes.  Such model predictions have important implications which we explore in the following section.

\section{Discussion}
\label{sec:discussion}

In this letter, we present predictions from the hydrodynamic cosmological \textsc{simba} simulations showing there is a dramatic increase in the molecular-gas mass to dust mass ratios, $\delta_{\mathrm{GDR}}$, when star formation slows and falls below log(sSFR)$\lesssim$-10 yr$^{-1}$.  The scatter in $\delta_{\mathrm{GDR}}$ spans  $>$4 orders of magnitude, with some quiescent galaxies having ``normal'' ratios whereas most others have exotic ratios. 

The wide range of predicted $\delta_{\mathrm{GDR}}$ values within the simulations cannot be explained by variations in metallicity as almost all model galaxies at low sSFR have solar or super-solar metallicities. Instead, it is likely that there exists rapid dust destruction in some galaxies shutting down star formation -- beyond dust consumed in the major episode of star formation itself.  The \textsc{simba} simulation includes three processes that destroy dust in low sSFR galaxies: (1) thermal sputtering by hot electrons, and destruction by (2) astration (the incorporation of dust into a stellar interior - a star particle here - during star formation) prior to the quenching (which is mainly due to supermassive black hole feedback) combined with (3) unresolved SN shocks.  All three play a role here (e.g., Figure~\ref{fig:test}), driving a rapid decrease in the predicted dust masses when star formation quenches.  Thermal sputtering in particular ramps up as gas density and/or temperature increases, and is invoked to explain high $\delta_{\mathrm{GDR}}$ values in nearby post-starburst galaxies \citep[e.g.,][]{Smercina18}.  

Interestingly, the first cold dust continuum detections of low sSFR galaxies outside the local universe remain unresolved despite the extreme extended stellar light profiles 
resulting from strong gravitational lensing \citep{Whitaker21}. 
Though only a sample of two, these data suggest high molecular gas and dust surface densities. This has also been found in star-forming galaxies at similar redshifts \citep[e.g.,][]{Tadaki17} and nearby post-starburst galaxies \citep{Smercina21}. Star formation may be suppressed by significant turbulent heating under these conditions, as internal turbulent pressure is proportional to gas surface density \citep{Smercina21}. Unfortunately, the simulations cannot resolve effects down to physical scales of the individual clouds.  Empirical confirmation of molecular-gas surface densities in significant samples of quiescent galaxies would corroborate this idea.

\begin{figure}[t]
\centering
\includegraphics[width=0.95\linewidth]{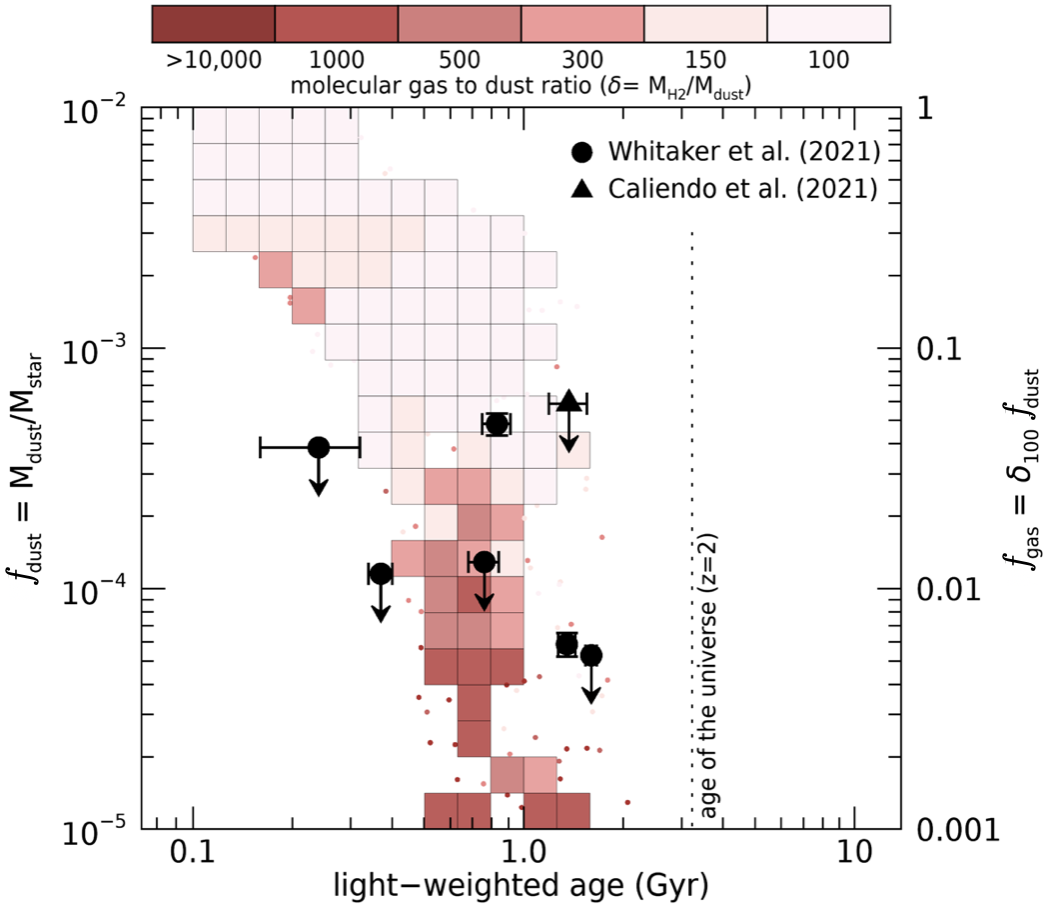}
\caption{At $z$=2, there exists a large scatter in $f_{\mathrm{dust}}$ at light-weighted ages older than 500 Myr. More observations of dust in quiescent galaxies, even deeper than those augmented by strong lensing \citep[black][]{Caliendo21, Whitaker21}, are required to test the predicted exotic molecular-gas to dust ratios and the steep turnover for light-weighted ages $>$500 Myr.}
\label{fig:mocks}
\end{figure}

Chemical evolution models presented in \citet{Remy14} predict that galaxies with the shortest star formation 
timescales of $\lesssim$0.5 Gyr sustain high $\delta_{\mathrm{GDR}}$ ratios for the longest time periods, up until they 
reach high metallicity, at which point $\delta_{\mathrm{GDR}}$ drops to more typical values on average.  The majority of quiescent galaxies in the \textsc{simba} simulation do not experience such a late-stage drop in $\delta_{\mathrm{GDR}}$ (e.g., Figure~\ref{fig:hist}) despite having high metallicity. This effect in the \citet{Remy14} models is the result of dust growth by the accretion of metals, but the dust mass in \textsc{simba} is too small to recover from depletion.

It is also possible for dust production to increase once again as the stellar populations age.  A common process to replenish dust reservoirs is via AGB stars. This phase of stellar evolution occurs when stars are around $\sim1$ Gyr, similar to the ages of quiescent galaxies at cosmic noon. 
AGB stars produce copious amounts of dust that could replenish dust reservoirs. However, nearby studies of quiescent
galaxies find that $f_{\mathrm{dust}}$ \citep{Michalowski19} and $f_{\mathrm{H2}}$ \citep{French15} both 
correlate with the age of the stellar populations, with older galaxies having lower dust and molecular-gas masses. In Figure~\ref{fig:mocks}, 
we consider the trend within the simulations at $z$=2 between $f_{\mathrm{dust}}$ and the light-weighted age, color-coded by $\delta_{\mathrm{GDR}}$.  The predicted trend in the high redshift universe is not gradual. There is a dramatic turnover for ages $>$500 Myr. We defer a more detailed analysis using star formation histories to discern between typical and exotic $\delta_{\mathrm{GDR}}$ in quenched galaxies to future work.  

The exotic molecular-gas to dust mass ratios predicted by \textsc{simba} call into question the efficiency with which one can use the observed dust continuum as an indirect measure of the interstellar medium at low sSFRs. To develop a more complete understanding of the efficacy of this technique, we must secure detections of both CO emission and dust continuum (ideally in multiple bands)
for the same galaxies. Uncertainties in the temperature and composition of dust in quiescent galaxies introduce important systematics when estimating the molecular-gas mass from a single dust continuum measurement \citep{Liang19,Privon18}.  Whereas such observations are costly with our current generation
of telescopes, the boost from gravitational lensing for spectacular sources 
offers a pathway forward in the near term for small samples.  Upcoming upgrades to existing facilities will also be fruitful: the improved sensitivity 
and wide field of view of the Toltec instrument on the Large Millimeter Telescope will enable similarly sensitive dust continuum studies of 
mass-representative quiescent samples at high redshift, and the spectroscopic capabilities predicted for the 
next generation Very Large Array will explore new parameter space for CO in these sources. Until then, the simulations paint a grim picture for the future utility of dust continuum for most low sSFR galaxies.

Predictions from \textsc{simba} encompass an extreme range of $\delta_{\mathrm{GDR}}$ ratios that are not yet observed. 
Owing to the scatter in these predicted ratios, the models do not elucidate a clear physical observable (i.e., stellar mass, surface density, sSFR, etc) to distinguish the expected $\delta_{\mathrm{GDR}}$ \citep{Li19}. But as a whole these predictions can be empirically tested and validated; future deeper CO observations and/or larger statistical dust continuum samples will help improve our understanding of the redshift evolution and behavior of $\delta_{\mathrm{GDR}}$ at low sSFR.  While current dust continuum sensitivity limits preclude confirmation of the low dust fractions predicted in Figure~\ref{fig:delta}, deeper CO measurements have the power to rule out these exotic ratios. By also measuring the spatial distributions of the molecular-gas, we can confirm the compact nature. If the model outcomes cannot be verified, the assumptions about dust processes must be revisited. There exists a beautiful synergy between these future observations and the model outcomes that will guide both observers and theorists alike.

\section*{acknowledgements}
We thank the anonymous referee for constructive feedback that helped to improve this manuscript.  
K.E.W. wishes to acknowledge funding from 
the Alfred P. Sloan Foundation, HST-GO-14622, and HST-GO-15663.  D.N. acknowledges funding from the National Science Foundation via AST-1908137. C.C.W. acknowledges support from  NIRCam Development Contract NAS5-02105 from NASA Goddard Space Flight Center to the University of Arizona.  Q.L. was funded from the National Science Foundation via AST-1909153.
S.T. acknowledges support from the ERC Consolidator Grant funding scheme
(project ConTExt, grant No.648179). The Cosmic Dawn Center is funded by the Danish National Research Foundation.  The simulations presented here were performed on the HiPerGator supercomputing facility at the University of Florida.

\end{document}